\documentclass[aps,prb,twocolumn,groupedaddress,showpacs,amsmath,amssymb,amsfonts]{revtex4}

\usepackage{graphicx}

\usepackage{color}
\definecolor{gray}{rgb}{0.7,0.7,0.7}

\renewcommand{\v}[1]{\ensuremath{\mathbf{#1}}} % for vectors
\renewcommand{\k}{\v{k}}
\newcommand{\avg}[1]{\left< #1 \right>} % for average
\newcommand{\abs}[1]{\left\vert #1 \right\vert} % for absolute value

\begin{document}

\title{Spontaneous dimerization in the spin-1 bilinear-biquadratic Heisenberg model on a honeycomb lattice}

\author{Yu-Wen Lee}
\email{ywlee@thu.edu.tw} %
\affiliation{Department of Physics, Tunghai University, Taichung
40704, Taiwan}

\author{Min-Fong Yang}
%\email{mfyang@thu.edu.tw} %
\affiliation{Department of Physics, Tunghai University, Taichung
40704, Taiwan}

\date{\today}

\begin{abstract}
Within the linear flavor-wave theory, we show that, due to the
quantum order-by-disorder mechanism, the spin-1
bilinear-biquadratic Heisenberg model on a honeycomb lattice can
spontaneously develop a columnar dimer order with a non-bipartite
structure. The low-lying excitations above this novel ground state
form several flat bands separated by nonzero energy gaps. Our
results suggest that the quantum phase transition separating this
dimerized phase with the nearby N\'{e}el-order phase may be of
first order.
\end{abstract}


\maketitle

%%%%%%%%%%%%%%%%%%%%%%%%%%%%%%%%%%%%%%%%%%%%%%%%%%%%%%%%%%%%%%%%%%
%{\it Introduction.}--- %

Quantum spin systems on various kinds of lattices have provided a
wide playground for the search of novel quantum states and quantum
phase transitions. It has been proposed that, in some spin-1/2
systems on a square lattice, quantum paramagnetic phases with
valence-bond-solid order can emerge and their transitions to the
antiferromagnetic N\'{e}el states may belong to the deconfined
quantum criticality.~\cite{Senthil04-1,Senthil04-2} On a honeycomb
lattice, even the more fascinating quantum spin-liquid state has
recently been proposed and even observed in the half-filled Hubbard model and
related frustrated spin-1/2 Heisenberg
models.~\cite{Meng10,FaWang,Clark11}

Rich physics is also expected in quantum spin systems with larger
spin moments. Prominent examples include the spin-1
bilinear-biquadratic (BLBQ) model described by the Hamiltonian
\begin{equation}
H=\sum_{\avg{i, j}}\left[J\left(\v{S}_i\cdot\v{S}_j\right) +K
\left(\v{S}_i\cdot\v{S}_j\right)^2\right]\,  \label{Hamiltonian}
\end{equation}
with a bilinear exchange coupling $J\equiv\cos\theta$ and a
biquadratic exchange coupling $K\equiv\sin\theta$, where the
parameter $\theta$ controls the ratio of these two couplings. It
has been shown that quantum phases with either collinear or
noncollinear nematic order can appear in this spin
system.~\cite{Papanicolaou88,Penc_Laeuchli} In these nematic
phases, dipolar spin order parameters vanish, $\langle \mathbf{S}
\rangle =0$, but spin rotation symmetry is spontaneously broken
due to nonzero quadrupolar spin expectation values, $\langle
{S^\alpha S^\beta + S^\beta S^\alpha } \rangle\neq0$ ($\alpha$,
$\beta=x$, $y$, $z$). Here we focus on the parameter region with
$\pi/4\leq\theta<\pi/2$ (i.e., $K\geq J>0$), where the
noncollinear nematic states described by mutually orthogonal
nematic directors are expected at the mean-field
level.~\cite{Papanicolaou88,Penc_Laeuchli} When the BLBQ model is
defined on a triangular lattice, in which each site has 6
neighbors, the ground state is found to posses a three-sublattice
nematic order.~\cite{Tsunetsugu06,Lauchli06} The nematic directors
on the three sublattices A, B, and C of the triangular
lattice are orthogonal to each other (say, along the $x$, $y$, and
$z$ directions, respectively).
On the other hand, if this model is placed on a square lattice,
because of the smaller coordination number for this lattice
structure, there are not enough constraints to uniquely determine
a set of mutually orthogonal nematic directors. Therefore, the
expected noncollinear nematic state will form a highly degenerate
ground-state manifold.
Usually, such macroscopic degeneracy at the variational level can
be lifted by quantum effects, and a unique ground state will be
selected by the order-by-disorder mechanism. Following this
reasoning, the ground state for the BLBQ model with $J=K$ on a
square lattice has been analyzed recently by means of a
semiclassical flavor-wave theory and exact
diagonalizations.~\cite{Toth10} It is found that instead of the
naive two-sublattice state, the ground state develops an
unexpected three-sublattice long-range order, even though
this system is defined on a bipartite lattice and with only
nearest-neighbor interactions.

Since the honeycomb lattice has an even smaller coordination
number, one may wonder whether it can support ground states quite
different from those on a square lattice. A possible candidate of
the quantum phase for the BLBQ model on a honeycomb lattice with
$\pi/4\leq\theta<\pi/2$ (i.e., $K\geq J>0$) has been
proposed.~\cite{Zhao2011} By employing the tensor renormalization
group method~\cite{TRG-1,TRG-2} under the assumption that the
ground-state wave function can be described by a periodic tensor
network with elementary hexagon as the unit cell, the ground state
is found to have plaquette valance-bond-solid (PVBS) order. This
PVBS state breaks the lattice translational symmetry but preserves
the spin SU(2) symmetry.
However, due to the assumed periodicity for their variational
states, the possibilities for the ground states to display more
complicated structures are missed in their exploration.
Hence their results may not be conclusive.

\begin{figure}[tb]
\includegraphics[clip,width=3.2in]{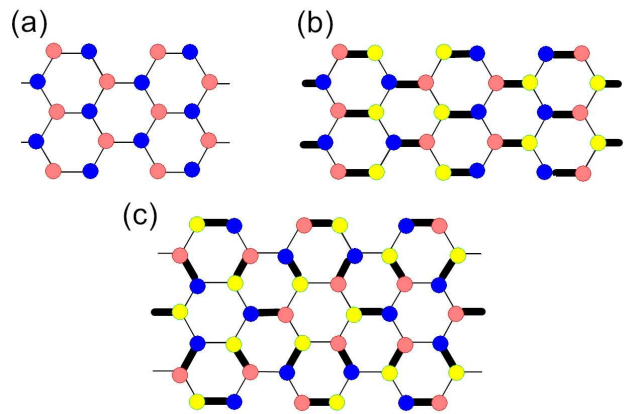}
\caption{(Color online) %
Schematic representation of three possible candidates for the
ground state: %
(a) two-sublattice state and
(b) staggered dimer state, both of which have higher LFW energies; %
(c) columnar dimer state selected by quantum fluctuations within the LFW theory. %
Here the three mutually orthogonal vectors $\mathbf{d}_i$ in the
mean-field analysis are denoted by the unit vectors along the $x$,
$y$, and $z$ directions, and are associated with different colors.
Strong bonds with the lowest zero-point energy are denoted by
thick lines.
The remaining weak bonds are depicted by thin lines.}
\label{fig:pattern}
\end{figure}

In the present work, we determine the nature of the ground state
of the BLBQ model in Eq.~\eqref{Hamiltonian} with $K\geq J>0$ on a
honeycomb lattice using the linear flavor-wave (LFW)
theory.~\cite{Papanicolaou88,Penc_Laeuchli,Chubukov90,Joshi99}
Previous studies on the cases of the square lattice have proved the
validity and success of this approach.~\cite{Toth10,Corboz2011} In
Ref.~\onlinecite{Toth10}, the conclusion of the LFW analysis for
the square-lattice case was shown to be supported by the exact
diagonalization calculations for systems of finite
sizes.~\cite{Toth10}
Besides, the results within the LFW theory for the SU(4)
Heisenberg model provide useful guides for numerical methods to
uncover the intricate dimerized structure and color ordering of
the ground state.~\cite{Corboz2011}
For the present case of the spin-1 BLBQ model on a honeycomb
lattice, following the reasoning of these works, we find that the
ground states exhibit exotic columnar dimer order [see
Fig.~\ref{fig:pattern}(c)] and break spontaneously both the spin
SU(2) symmetry and the lattice translation symmetry. Our results
arise from the order-by-disorder mechanism, where the ground
states are selected among the degenerate manifold of the
noncollinear nematic states by minimizing the zero-point energies
of quantum fluctuations.
We note that the unit cell of the resulting ground-state pattern
is quite large (consisting of 18 lattice sites), while the
underlying lattice is bipartite.
Within our LFW analysis, the flavor-wave excitations in the
present systems are found to be localized, and the low-lying modes
form several flat bands separated by nonzero energy gaps. This
observation is rather different from the usual cases with
spontaneous SU(2) symmetry breaking.
Moreover, it is found that the gapful ground state with columnar
dimer order remains even in the limit of $J=K$ (or
$\theta=\pi/4$), at which a direct transition to the nearby
antiferromagnetic phase for $J>K$ is
anticipated.~\cite{Papanicolaou88,Penc_Laeuchli} This suggests
that the quantum phase transition at $J=K$, which separates these
two phases with distinct types of long-range order, may be of
first order.
The implication of our results on generalized models is discussed
at the end of this Rapid Communication.

%%%%%%%%%%%%%%%%%%%%%%%%%%%%%%%%%%%%%%%%%%%%%%%%%%%%%%%%%%%%%%%%%%
%{\it The flavor-wave analysis.}--- %

The LFW theory starts from representing the model in
Eq.~\eqref{Hamiltonian} in terms of three-flavor Schwinger bosons
$a_{i,\alpha}$ under the local constraint $\sum_{\alpha}
a_{i,\alpha}^\dagger a_{i,\alpha}=1$,
\begin{equation}\label{boson_Hamil}
H=\sum_{\avg{i,j}} \left[ J \chi_{ij}^\dagger \chi_{ij} + (K-J)
\Delta_{ij}^\dagger \Delta_{ij} + (K-J) \right] \; .
\end{equation}
Here we define two bond operators, $\chi_{ij}=\sum_{\alpha}
a_{i,\alpha}^\dagger a_{i,\alpha}$ and $\Delta_{ij}=\sum_{\alpha}
a_{i,\alpha}a_{j,\alpha}$. The Schwinger bosons
$a_{i,\alpha}^\dagger$ (with $\alpha=x$, $y$, $z$) create
three time-reversal-invariant local basis states, %
$|x\rangle = \frac{1}{\sqrt{2}}(|s_z=1\rangle - |s_z=-1\rangle)$, %
$|y\rangle = \frac{i}{\sqrt{2}}(|s_z=1\rangle + |s_z=-1\rangle)$, and %
$|z\rangle = |s_z=0\rangle$. %
In terms of these bosons, the spin operators become %
$S_{i,\alpha}=-i\sum_{\beta,\gamma}\epsilon_{\alpha\beta\gamma}a_{i,\beta}^\dagger
a_{i,\gamma}$.
The first step is the mean-field analysis based on a
site-factorized variational wave function. At this mean-field
level, the Schwinger-boson operators $\mathbf{a}_i\equiv(a_{i,x}$,
$a_{i,y}$, $a_{i,z})$ are replaced by a (complex) three-component
vector $\mathbf{d}_i$, and the energy of the nearest-neighbor bond
$\avg{i,j}$ is minimal when the two vectors $\mathbf{d}_i$ and
$\mathbf{d}_j$ are mutually
orthogonal.~\cite{Papanicolaou88,Penc_Laeuchli} The mean-field
ground state configuration is highly degenerate on a honeycomb
lattice, and as can be seen from Eq.~\eqref{boson_Hamil}, the
associated mean-field energy per bond is $(K-J)$. This macroscopic
degeneracy can be lifted when quantum fluctuations above each
mean-field state are included. Within the LFW analysis, the
leading quantum corrections to the mean-field energy of the
considered variational state come from the zero-point energy of
the LFW Hamiltonian. Therefore, the configuration with the lowest
zero-point energy will be picked out as the true ground state.
This is in essence a quantum order-by-disorder selection
mechanism.

%------------------  figure  --------------------------------
%------------------  figure  --------------------------------

The derivation of the LFW Hamiltonian proceeds as follows. For a
given configuration of the variational state, when the two
classical vectors on a nearest-neighbor bond $\langle i,j\rangle$
are, say, $\mathbf{d}_i=\hat{x}$ and $\mathbf{d}_j=\hat{y}$ (named
as the $XY$ bond in the following), we approximate semiclassically
their Schwinger-boson operators by $\mathbf{a}_i \simeq (1$,
$a_{i,y}$, $a_{i,z})$ and $\mathbf{a}_j \simeq (a_{j,x}$, 1,
$a_{j,z})$. The operators $a_{i,y}$ and $a_{i,z}$ on site $i$
($a_{j,x}$ and $a_{j,z}$ on site $j$) describe the quantum
fluctuations around the classical vectors, and they play the role
of the Holstein-Primakoff bosons in the usual spin-wave theory.
Therefore, up to the linear order in these operators, the bond
operators become
\begin{equation}
\chi_{ij} \simeq a_{i,y}^\dagger + a_{j,x} \, , \quad\quad
\Delta_{ij}\simeq a_{i,y}+a_{j,x} \, . \label{bond_operators}
\end{equation}
Substituting them into the Hamiltonian in Eq.~\eqref{boson_Hamil},
we obtain the desired quadratic LFW Hamiltonian.

We note that the LFW Hamiltonian in general consists of a
sum of independent parts that describe the motion of bosons on
certain connected clusters. For example, for a given
nearest-neighbor bond (say, the $XY$ bond), when all of the
$\mathbf{d}$ vectors of its surrounding sites are orthogonal (say,
$\mathbf{d}_i=\hat{z}$) to \emph{both} $\mathbf{d}$ vectors on
that bond, the 2 bosons (say, the $x$ and $y$ components of the
Schwinger bosons) within that bond cannot move to neighboring
sites. That is, the motion of these bosons becomes decoupled from
the surrounding of that bond, and it can be described by a 2-site
Hamiltonian. In the following, such 2-site clusters are dubbed as
strong bonds and denoted by thick lines in
Figs.~\ref{fig:pattern}(b) and \ref{fig:pattern}(c). On the other hand, the
remaining bonds depicted by thin lines can be linked to from
larger cluster and they are termed as weak bonds. As discussed
below, we find that the zero-point energy (and therefore the
ground-state energy) is minimized for those strong bonds.
Therefore, we expect that the states with the lowest zero-point
energy should be the ones that contain as many strong bonds as
possible. Such a ``maximum strong bond rule," which has been
noted in other context,~\cite{Corboz2011} forms our main guiding
principle in searching for possible candidates of the ground
state.
Based on this observation, in addition to the two-sublattice state
[Fig.~\ref{fig:pattern}(a)], which is the naive ground-state
configuration on the present bipartite lattice, we consider two
more configurations. They are the staggered dimer state and the
columnar dimer state [Figs.~\ref{fig:pattern}(b) and Figs.~\ref{fig:pattern}(c),
respectively], both of which contain the maximum number of strong
bonds per elementary hexagon.
Nevertheless, as discussed in the paragraph below
Eq.~(\ref{3_sublattice}), the zero-point energy for weak bonds in
the columnar dimer state is lower than that in the staggered dimer
state. Thus the columnar dimer state in Fig.~\ref{fig:pattern}(c)
with a more complicated structure is selected by the zero-point
quantum fluctuations within the LFW theory.

Now we begin to derive the explicit expressions of the zero-point
energies for the three configurations shown in
Fig.~\ref{fig:pattern}.
For the case of the two-sublattice state formed by $XY$ bonds
only, the $z$ components of the Schwinger bosons play no role and the LFW
Hamiltonian reduces to an effective model for the two-component
boson mixture on a single connected cluster of the whole honeycomb
lattice. By means of the Bogoliubov transformation, the excitation
spectrum of the resulting quadratic bosonic Hamiltonian can be
easily obtained. This results in the following ground-state energy
per site (i.e., the sum of the mean-field energy density and the
zero-point contribution),
$E_\textrm{G}=\frac{1}{N}\sum_\k (\lambda_{1\k}+\lambda_{2\k})$.
Here $N$ is the number of lattice sites, $\k$ runs over the first
Brillouin zone of the honeycomb lattice, and
$\lambda_{1,2}(\k)=\sqrt{\left(\varepsilon_0\pm
\abs{\Delta_{2\k}}\right)^2-\abs{\Delta_{1\k}}^2}$
with $\varepsilon_0=\frac{z}{2}K$,
$\Delta_{1\k}=\frac{z}{2}J\gamma_\k$, and
$\Delta_{2\k}=\frac{z}{2}(K-J)\gamma_\k$. %
The coordination number $z=3$ for the honeycomb lattice, and
$\gamma_\k=\frac{1}{z}\sum_\delta e^{i\k\cdot\delta}$, where
$\delta$ runs over the three nearest-neighbor vectors,
$\delta_{1}=(1,0)$, $\delta_2=(-\frac{1}{2},\frac{\sqrt{3}}{2})$,
and $\delta_3=(-\frac{1}{2},-\frac{\sqrt{3}}{2})$.

Unlike the two-sublattice state, both the staggered dimer state
and the columnar dimer state have strong bonds. Besides, their
weak bonds form many one-dimensional zig-zag chains or 6-site
clusters (hexagonal loops) as shown in Fig.~\ref{fig:pattern}(b)
and Fig.~\ref{fig:pattern}(c), respectively. Since the bosons on
these weak bonds are decoupled from those on the strong bonds,
these bosons can be described by an effective Hamiltonian on a
one-dimensional chain of 2$N_c$ bonds, where $N_c\to\infty$ and
$N_c=3$ for the staggered dimer state and the columnar dimer
state, respectively. That is, the flavor-wave excitations are well
localized within those chains or clusters.
Because the LFW Hamiltonians on clusters of different sizes give
distinct contributions to the zero-point energy, they need to be
considered separately.
For a given strong bond, by diagonalizing the corresponding
2-site LFW Hamiltonian, its ground-state energy can be found as
\begin{equation}
E_\textrm{s-bond}= \sqrt{K(K-J)} \, . \label{E_strong_bond}
\end{equation}
Similarly, for excitations localized within a closed chain of
$2N_c$ bonds, diagonalization of the corresponding one-dimensional
Hamiltonian results in the following expression for the
ground-state energy per weak bond,
\begin{equation}
E_\textrm{w-bond}(N_c)
=\frac{1}{2N_c}\sum_k\left[\tilde{\lambda}_1(k)+\tilde{\lambda}_2(k)\right]
\, . \label{E_weak_bond}
\end{equation}
Here the one-dimensional momentum $k=\frac{2\pi n}{N_c}$ with
$n=1,\ldots,N_c$,
$\tilde{\lambda}_{1,2}(k) =\sqrt{\left(\tilde{\varepsilon}_0 \pm
|\tilde{\Delta}_{2k}|\right)^2-|\tilde{\Delta}_{1k}|^2}$
with $\tilde{\varepsilon}_0=K$,
$|\tilde{\Delta}_{1k}|=J|\cos(\frac{k}{2})|$, and
$|\tilde{\Delta}_{2k}|=(K-J)|\cos(\frac{k}{2})|$.
As seen from Figs.~\ref{fig:pattern}(b) and \ref{fig:pattern}(c), in both the
staggered dimer state and the columnar dimer state, the three
nearest-neighbor bonds for each lattice site contain one strong
and two weak bonds. Therefore, from Eqs.~\eqref{E_strong_bond} and
\eqref{E_weak_bond}, the expression of the ground-state energy per
site for both cases becomes
\begin{equation}
E_\textrm{G}=\frac{1}{2}E_\textrm{s-bond} + E_\textrm{w-bond}(N_c)
\, . \label{3_sublattice}
\end{equation}
Thus the difference in energies between the staggered dimer state
and the columnar dimer state comes from the contribution of the
weak bonds. From Eq.~\eqref{E_weak_bond}, one can show that the
energy per weak bond is an increasing function of $N_c$. In other
words, the localized flavor-wave excitations residing on shorter
chains will give a smaller contribution in
Eq.~(\ref{3_sublattice}). As mentioned above, $N_c=3$ for the
columnar dimer state and $N_c\to\infty$ for the staggered dimer
state. Hence, we conclude that the former has a lower energy.
Note that on a honeycomb lattice, the shortest closed loop of
weak bonds is nothing but the hexagonal one with $N_c=3$. Thus we
conclude that the columnar dimer state should be the true ground
state among the degenerate manifold.

%------------------  figure  --------------------------------
\begin{figure}
\includegraphics[clip,width=3.2in]{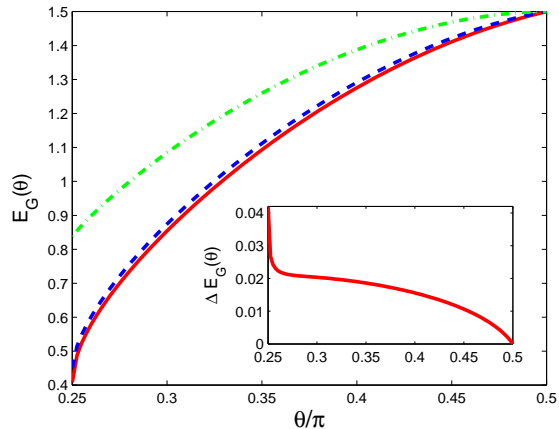}
\caption{(Color online) %
Ground state energy densities (per site) for the three states
shown in Fig.~\ref{fig:pattern} as a function of
$\theta=\tan^{-1}(K/J)$. The dash-dotted, dashed, and solid
lines are for the two-sublattice, the staggered dimer, and the
columnar dimer states, respectively.
Inset: Energy difference (per site) $\Delta E_G\equiv
E_{G,\textrm{staggered dimer}}-E_{G,\textrm{column dimer}}$
between the staggered dimer and the columnar dimer states as a
function of $\theta$.
} \label{fig:stripe-dimer}
\end{figure}
%------------------  figure  --------------------------------

The ground-state energy densities for the three types of states
are presented in Fig.~\ref{fig:stripe-dimer}.
We find that their energies become degenerate and approach to
$3/2$ in the $\theta\to\pi/2$ limit. This comes from the fact that
the mean-field state becomes an exact eigenstate at $\theta=\pi/2$
(or $J=0$ and $K=1$) and quantum fluctuations play no role here.
Thus the mean-field energy per site
$E_\textrm{G}=\frac{z}{2}(K-J)=3/2$ is nothing but the exact
energy eigenvalue.
Moreover, our Fig.~\ref{fig:stripe-dimer} shows that the columnar
dimer state does have the lowest energy in the whole parameter
regime $\pi/4\leq\theta=\tan^{-1}(K/J)<\pi/2$.
We note that the PVBS states proposed in
Ref.~\onlinecite{Zhao2011} lie outside the degenerate manifold of
the noncollinear nematic states and thus are not considered within
the employed approach. However, we can compare their variational
energies with those discussed in the present work. It is found that
the ground-state energies of our columnar dimer states are also
lower than those of the PVBS states. For instance, at the SU(3)
point with $\theta=\pi/4$ (or $J=K=1/\sqrt{2}$), our values of
$E_G$ for the two-sublattice state, the staggered dimer state, and
the columnar dimer state are $0.8381$, $\sqrt{2}/\pi\;(\simeq
0.4502)$, and $1/\sqrt{6}\;(\simeq 0.4082)$, respectively.
However, the energy density at $\theta=\pi/4$ for the PVBS state
is about 0.536, which is higher than the values for both of the
dimerized states. Thus our calculations show that the peculiar
dimerized ground states that spontaneously break both spatial and
spin SU(2) symmetries can in fact be energetically more favorable.
We should stress that since the energies obtained within the LFW
theory are not variational, the above comparison in energy may not
be used to exclude the PVBS states as the true ground state.
Nevertheless, the present investigation indicates that our
dimerized states should at least be possible candidates.

To provide guides for future numerical investigations, some
comments are discussed below.
People usually perform exact diagonalization calculations for
systems of finite sizes to obtain unbiased results and then
provide numerical verification for analytical proposals. Since the
unit cell of our columnar dimer state contains 18 lattice sites,
large systems with sizes of several unit cells should be
considered in order to deliver meaningful results complementary to
our present work.
But the required computational resources may make such
calculations unlikely. In this regard, numerical variational
approaches, such as the variational Monte Carlo method and the
tensor network method, may be more promising, because their size
limitation is less severe.
Even in this case, our results show that one has to
use the variational states compatible with the spatial structure
of our columnar dimer state to determine convincingly
the true ground state.

In addition to determining the nature of the ground state within
parameter regime $\pi/4<\theta<\pi/2$, our results also help to
characterize the types of phase transitions out of this phase. It
is known that at the mean-field level, a phase transition to the
nearby ferromagnetic phase occurs at
$\theta=\pi/2$.~\cite{Papanicolaou88,Penc_Laeuchli} At this value
of $\theta$, we find that the energy of the columnar dimer state
($E_\textrm{G}=3/2$) is identical to that of the ferromagnetic
state [see Eq.~\eqref{Hamiltonian} with $\v{S}_i=1$]. Because
these two states belong to different subspaces of the total spin,
this agreement in energy indicates a level crossing and thus a
first-order transition at $\theta=\pi/2$.
On the other hand, the mean-field theory predicts a direct
transition at $\theta=\pi/4$ from the noncollinear nematic to the
antiferromagnetic phases.~\cite{Papanicolaou88,Penc_Laeuchli} In
our work, the effects of quantum fluctuations are taken into
account under the LFW analysis. We find that there exist only the
localized flavor-wave excitations above the columnar dimer state
and these excitations form flat bands separated by nonzero energy
gaps. Moreover, the gapful ground state with nonvanishing columnar
dimer order persists even in the limit of $\theta=\pi/4$ (or
$J=K$). This suggests that the quantum phase transition at $J=K$,
which separates these two phases with distinct types of long-range
order, should be of first order.

Finally, the implication of our results on generalized models is
discussed. Unlike its counterparts on other lattices with larger
coordination numbers $z$ [i.e., the three-sublattice
antiferronematic state on a triangular
lattice~\cite{Tsunetsugu06,Lauchli06} ($z=6$) and the
three-sublattice stripelike state on a square
lattice~\cite{Toth10} ($z=4$)], the columnar dimer state on a
honeycomb lattice ($z=3$) has a solidlike pattern and supports
only localized flavor-wave excitations within the LFW analysis. In
addition, a conclusion similar to ours is reached for an
SU(4)-symmetric spin model with 4 states in the local Hilbert
space.~\cite{Corboz2011} From all these findings, one may conclude
that for generalized models with more local states [say, the
SO($n$) bilinear-biquadratic model~\cite{Tu08,Alet11,Tu11}] and/or
on two-dimensional lattices with smaller $z$, ground states with
solid-like structures will be stabilized due to the same
order-by-disorder mechanism. While this observation is based on
the LFW analysis, we believe that the qualitative picture will not
be modified even if higher quantum corrections are included.

To summarize, by employing the LFW theory, we show that the
columnar dimer state is the ground state of the spin-1 BLBQ model
on a honeycomb lattice with $\pi/4<\theta<\pi/2$.
We stress that our investigations are not of purely academic
interest. In fact, at $\theta=\pi/4$ (i.e., $K=J>0$), the present
model possesses an enlarged SU(3) symmetry and can be considered
as an effective model for the Mott-insulating state of
three-flavor cold fermions with one particle per lattice
site.~\cite{Toth10} Therefore, our conclusions at this SU(3) point
may be examined experimentally for such cold fermions on a
honeycomb optical lattice.~\cite{Optical}

%%%%%%%%%%%%%%%%%%%%%%%%%%%%%%%%%%%%%%%%%%%%%%%%%%%%%%%%%%%%%%%%%%
%\begin{acknowledgments}
Y.-W.L. and M.-F.Y. acknowledge the support from the National Science
Council of Taiwan under Grants No. NSC 99-2112-M-029-004-MY3 and No. NSC
99-2112-M-029-003-MY3, respectively.
%\end{acknowledgments}
%%%%%%%%%%%%%%%%%%%%%%%%%%%%%%%%%%%%%%%%%%%%%%%%%%%%%%%%%%%%%%%%%%

\end{document}